\documentclass[preprint, preprintnumbers,amsmath,amssymb]{revtex4}
\usepackage{graphicx}% Include figure files
\usepackage{dcolumn}% Align table columns on decimal point
\usepackage{bm}% bold math

\def\be{\begin{equation}}
\def\ee{\end{equation}}
\def\bea{\begin{eqnarray}}
\def\eea{\end{eqnarray}}

\begin{document}

\title{ Isospin Chemical potential and temperature effects in the Linear Sigma Model }

\author{M.L. El-Sheikh}
\email{mlelshei@uc.cl} \affiliation{Facultad de F\'\i sica,
Pontificia Universidad Cat\'olica de Chile,\\ Casilla 306, Santiago
22, Chile.}
\author{M. Loewe}
\email{mloewe@fis.puc.cl} \affiliation{Facultad de F\'\i sica,
Pontificia Universidad Cat\'olica de Chile,\\ Casilla 306, Santiago
22, Chile.}

\begin{abstract}
In this letter we explore the temperature and isospin chemical
potential ($\mu_I$) dependence of the vacuum structure and the meson
masses in the linear sigma model, at the one loop level. The sigma
meson mass grows steadily with temperature. This behavior does not
agree with previous mean field calculations. For the pion our
results show the same behavior as the mean field approach. The
stability of the vacuum has a very soft dependence on $\mu_I$ since
this is a second order effect in the tadpole diagrams.
\end{abstract}

\maketitle

\section{Introduction}

In this letter we address, once again, the question of how the
temperature (T) and the chemical potential, in our case isospin
chemical potential ($\mu_I$), affect the behavior of the linear
sigma model. The question is interesting, since this model, proposed
by Gell-Mann and Levy \cite{gell} represents perhaps the most simple
realization of QCD at low energies, being also renormalizable
\cite{renormalizacion}. Of course, there are other low energy
realization of QCD as, for example, the no-linear sigma model, and
its treatment according to Chiral perturbation theory
\cite{otrosapproaches}.

In both approaches for the low energy realization of QCD, the
fundamental questions concern the occurrence of phase transitions,
as, for example, restoration of Chiral symmetry, the presence of a
pion superfluid phase, etc, as well as the evolution of parameters
in the model with emphasis on mesonic masses and/or couplings
constants. The recent analysis in the frame of Chiral perturbation
theory for masses and phase transitions has been done in
\cite{villa}.

The novelty of this article is the analysis of the $\mu_I$ role,
together with T, in the context of these questions. Effects of pure
finite temperature have been discussed in the past many times
\cite{contreras} and \cite{larsen}. More recently, \cite{antifermi}
the Chiral symmetry restoration at finite T and finite baryon
density has been discussed. The chemical potential involved in their
analysis is the baryonic chemical potential. It is well known,
\cite{kogut}, that the dependence of the low energy effective
lagrange parameters on T and the chemical potential is different if
we consider the baryonic chemical potential or the isospin chemical
potential, which is related to the existence of an asymmetric charge
state of matter. The $\mu_I$, as well as the temperature dependence
in the linear sigma model have been discussed very recently in
\cite{nico}. Now, their treatment is based on the
Cornwall-Jackiw-Tomboulis (CJT), formalism, which allows them to
obtain the mass evolution from certain gap equations. An analysis of
finite isospin chemical potential as well as finite temperature
effects has been carried on in the frame of the Nambu-Jona- Lasinio
model, \cite{He}, where the authors also include a discussion of the
linear sigma model.

It is interesting, therefore, to compare the prediction of the one
loop correction, including $\mu_I$ and T, to the masses in the model
and compare the results with the approach in \cite{nico}. As we will
see the results for the pion mass evolution coincides almost
completely, been the behavior of the scalar meson sigma, however,
quite different in both approaches.

\vspace{3cm}

First, we will consider briefly the problem of chiral symmetry
restoration due to the tadpoles, which in fact depends only on T at
the one loop level. Then, we go on into the problem of the
determination of pion an sigma meson masses. The fermions, which in
the model correspond to the nucleons, are quite massive and stable.
In fact, regarding phase transitions only the lower mass states are
relevant.

%*****************************************************************
\section{The Linear Sigma Model and the evolution of the vacuum}
%******************************************************************

The linear sigma model is defined through the following lagrangian:

\begin{equation}
L=L_{0}+\epsilon L_{1}
\end{equation}
\begin{equation}
L_{0}=\overline{\psi}[i\gamma_{\mu}\partial^{\mu}-g(\sigma+\boldsymbol{\pi}\cdot\boldsymbol{\tau}\gamma_{5})]\psi
+\frac{1}{2}[(\partial_{\mu}\sigma)^{2}+(\partial_{\mu}\boldsymbol{\pi})^{2}]-\frac{\mu^{2}}{2}[\sigma^{2}+\boldsymbol{\pi}^{2}]
-\frac{\lambda^{2}}{4}[\sigma^{2}+\boldsymbol{\pi}^{2}]^{2}
\end{equation}
\begin{equation}
L_{1}=c\sigma
\end{equation}

Notice that we have a scalar field $\sigma$ and a vector of
isoscalars, the pions ($\boldsymbol{\pi}$). Notice also that due to
the presence of the explicit Chiral symmetry breaking term $L_1$, we
have $\langle0~|~\sigma|~0\rangle=v$, $v\neq0$. Shifting the
$\sigma$ field as $\sigma=s+v,$ were $\langle0~|~s|~0\rangle=0$, we
have a new lagrangian given by $L=L_{a}+L_{b}+L_{c}$:

\begin{equation}
L_{a}=\overline{\psi}[i\gamma\cdot\partial-m-g(s+i\boldsymbol{\pi}\cdot\boldsymbol{\tau}\gamma_{5})]\psi
+\frac{1}{2}[(\partial
\boldsymbol{\pi})^{2}-\mu_{\pi}^{2}\boldsymbol{\pi}^2]
\end{equation}
$$
+\frac{1}{2}[(\partial s)^{2}-\mu_{\sigma}^{2}
s^2]-\lambda^{2}vs(s^{2}+\boldsymbol{\pi}^{2})-\frac{\lambda^{2}}{4}(s^{2}+\boldsymbol{\pi}^{2})^{2}
$$
\begin{equation}
L_{b}=(\epsilon c-v\mu_{ \pi}^{2})s
\end{equation}
\begin{equation}
L_{c}=-\frac{\mu^{2}v^{2}}{2}-\frac{\lambda^{2}v^{4}}{4}+\epsilon
cv,
\end{equation}

\noindent where we have defined:

$$
m=gv
$$
\begin{equation}
\mu_{\pi}^{2}=\mu^{2}+\lambda^2v^2
\end{equation}
$$
\mu_{\sigma}^{2}=\mu^{2}+3\lambda^2v^2.
$$

\vspace{2cm}

We defined the charged pions as:

$$
\pi^{\pm}=\frac{\pi_{1}\mp i\pi_{2}}{\sqrt{2}}
$$

$$
\pi^{0}=\pi_{3}.
$$

The important object, in order to fix $v$, as function of T, is to
impose $\langle0~|~s|~0\rangle=0$. This means that the sum of
tadpoles must vanishes, i.e.:

\begin{equation}
\includegraphics[scale=0.4]{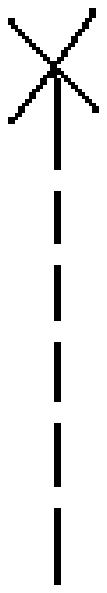}
+\frac{1}{2}
\includegraphics[scale=0.4]{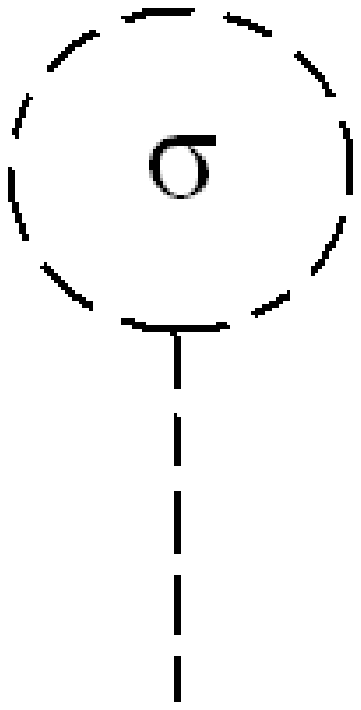}
+\frac{1}{2}
\includegraphics[scale=0.4]{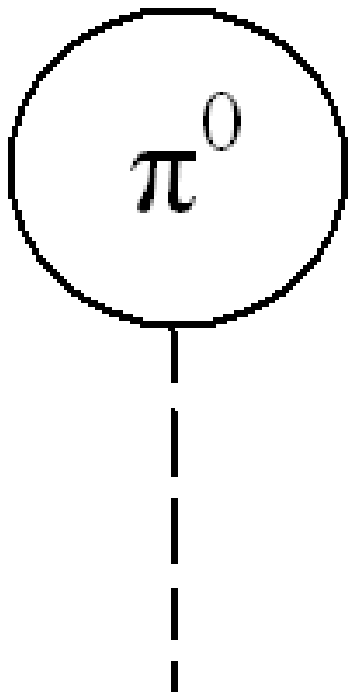}
+2
\includegraphics[scale=0.4]{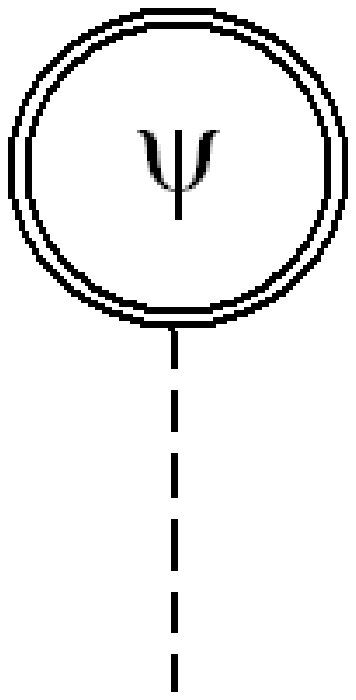}
=0
\end{equation}

This can be rewritten as:

\begin{equation}
i(\epsilon
c-v\mu_{\pi}^{2})+\frac{1}{2}T_{\sigma}+\frac{1}{2}T_{\pi}+2T_{\psi}=0
\end{equation}

In the previous equation, $T_{\sigma}$, $T_{\pi}$, $T_{\psi}$,
denote the corresponding tadpole corrections. Notice that we have
included symmetry factors.

If we neglect the chemical potential associated to the fermions, due
to their very high mass, the previous relation gives us an evolution
of $v=v(T)$. The $\sigma$ and $\pi_0$ fields are isoscalars, i.e.
there is no isospin chemical potential associated to them. The
influence of $\mu_I$ on the vacuum stability appears only in higher
loop tadpole corrections.

The thermal corrections for $v(T)$ are obtained by taking the
thermal part of the propagators in each tadpole.

Thermo Field Dynamics (TFD) is an appropriate formalism for dealing
with thermal loop corrections in the real time formalism. For one
loop corrections, the matrix propagators reduce to the well known
Dolan-Jackiw propagators. The propagator for a bosonic scalar field
is given by:

\begin{equation}
\Delta_{\beta}(k)=\frac{i}{k^{2}-m^{2}+i\epsilon}+2\pi
n_{B}(|k_{0}|)\delta(k^{2}-m^{2}),
\end{equation}
\vspace{1cm}

\noindent and for a fermion field, by

\begin{equation}
S_{\beta}(k)=\frac{i}{k\!\!\!\slash-m+i\epsilon}-2\pi
n_{F}(|k_{0}|)(k\!\!\!\slash+m)\delta(k^{2}-m^{2}).
\end{equation}

In the previous equations, $n_{B}$ and $n_{F}$, denote the well
known Bose-Einstein and the Fermi-Dirac statistical factors,
respectively.

$$n_{F,B}(z)=\frac{1}{exp(z)\pm1}.$$

The most natural way to introduce a chemical potential in field
theory is to consider it as a zero component of an external constant
gauge field ,\cite{Wel82} and \cite{Act85}. In this way, for example
for the scalar bosonic propagator we have now

\begin{equation}
D_{DJ}(k)_{+-}=\frac{i}{(k+\mu_{I}u)^{2}-m_{\pi}^{2}+i\epsilon}+2\pi
n_{B}(|k_{0}|)\delta((k+\mu_{I}u)^{2}-m_{\pi}^{2}).
\end{equation}

\vspace{1 cm}
 If we take only in to account the finite thermal
corrections we get

\begin{equation}
i(\epsilon
c-v\mu_{\pi}^{2})-iv(3\lambda^{2}I_{\sigma}+\lambda^{2}I_{\pi}-8g^{2}I_{\psi})=0,
\end{equation}

\noindent where,

\begin{equation}
I_{\sigma}=\int\frac{d^4k}{(2\pi)^4}\Delta_{\sigma}^{\beta}(k)=
\int\frac{d^4k}{(2\pi)^4}2\pi
n_{B}(|k_{0}|)\delta(k^2-m_{\sigma}^{2}),
\end{equation}

\begin{equation}
I_{\pi}=\int\frac{d^4k}{(2\pi)^4}\Delta_{\pi}^{\beta}(k)=
\int\frac{d^4k}{(2\pi)^4}2\pi n_{B}(|k_{0}|)\delta(k^2-m_{\pi}^{2}),
\end{equation}

\begin{equation}
I_{\psi}=\int\frac{d^4k}{(2\pi)^4}\Delta_{\psi}^{\beta}(k)=
(-1)\int\frac{d^4k}{(2\pi)^4}2\pi
n_{F}(|k_{0}|)\delta(k^2-m_{\psi}^{2}).
\end{equation}

\vspace{6cm}

The evolution of $v=v(T)$ is shown in Fig.1.We notice that $v$
vanishes only if $L_1$ is absent.

\begin{figure}[h]
\begin{center}
\includegraphics[scale=0.4]{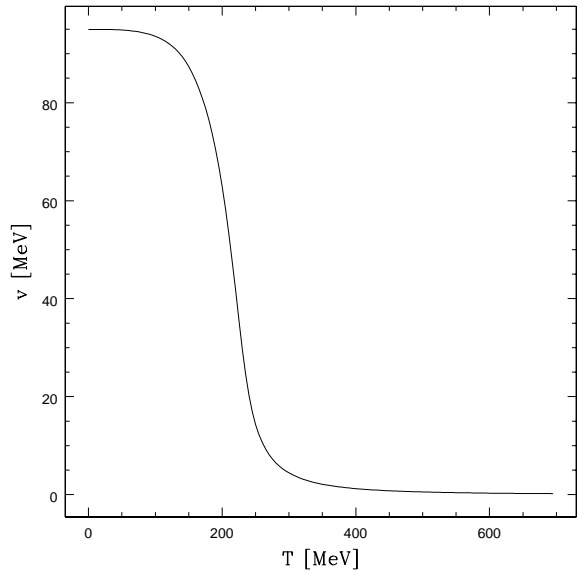}
\caption[Vacuum expectation value of the  $\sigma$ field, $v$(T)
]{Vacuum expectation value of the  $\sigma$ field,
$v$(T)}\label{vdet}
\end{center}
\end{figure}

\vspace{0.3 cm}

For the numerical purposes we will use, $m_{\pi}=139.6 ~~MeV$,
$m_{\sigma}=550 ~~MeV$, $m_{\psi}=900 ~~MeV$, i.e.
$m_{\pi}<m_{\sigma}<m_{\psi}$. Recently, \cite{bediaga}, the
existence of a scalar meson with $m\approx550~~ MeV$ has been
established.

An analytic treatment of $v(T)$ is only possible in the low or high
energy temperature regime. In Fig.1 we show the numerical results.
As a consequence of the thermal tadpoles, the tree level potential
energy of the model,

\begin{equation}
V(\sigma,\boldsymbol{\pi})=\frac{\lambda^{2}}{4}[\sigma^{2}+\boldsymbol{\pi}^{2}]^{2}+
\frac{\mu^{2}}{2}[\sigma^{2}+\boldsymbol{\pi}^{2}] -\epsilon c\sigma
\label{pot1}
\end{equation}

\noindent develops a non trivial temperature dependent minimum for
$\boldsymbol{\pi}=0$. In fact we get,

\begin{equation}
V(s,0)=\lambda^{2}vs^{3}+\frac{\lambda^{2}}{4}s^{4} -(\epsilon
c-v\mu_{ \pi}^{2})s,\label{pot3}
\end{equation}

\noindent where $v=v(T)$.

If we start from the Nambu-Goldstone phase, $\mu^2<0$, we can see
that for $0\leq T\lesssim 230 ~~MeV$ the potential has two minima as
it is shown in Fig.2,

\begin{figure}[h!]
\begin{center}
\includegraphics[scale=0.4]{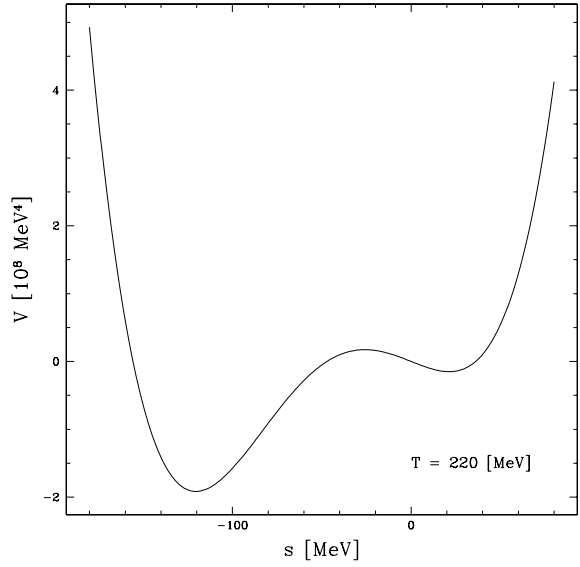}
\caption[Potential behavior for $T=220$ MeV]{Potential behavior for
$T=220$ MeV}\label{220}
\end{center}
\end{figure}

However, when $T=232~MeV$ the potential starts to develop only one
minimum. This fact, shown in Fig.3, represents the occurrence of a
phase transition from the Nambu-Golstone mode into de Wigner mode,
if we neglect the explicit chiral symmetry breaking term $L_1$, as
it is discussed in \cite{lee}.

\begin{figure}[h!]
\begin{center}
\includegraphics[scale=0.4]{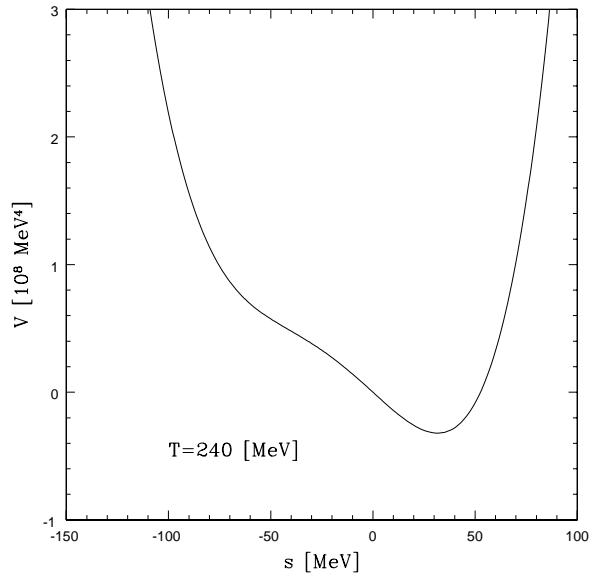}
\caption[Potential behavior for $T=240$ MeV]{Potential behavior for
$T=240$ MeV}\label{240}
\end{center}
\end{figure}

\pagebreak
%***************************************************************************************************
\section{Thermal and isospin chemical potential mass corrections}
%****************************************************************************************************

In order to find the mass corrections for the $\pi_0$ we need to
evaluate the diagrams shown in Fig.4,

\begin{figure}[h]
\begin{center}
\includegraphics[scale=0.5]{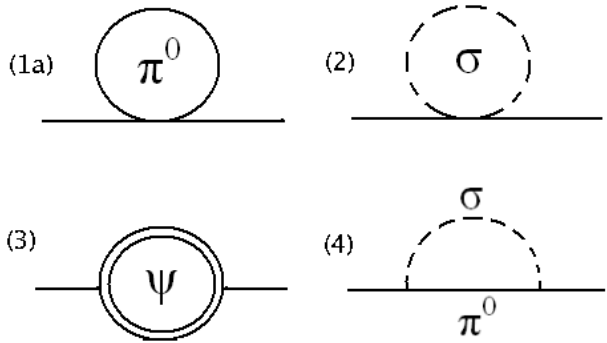}
\caption[Diagrams that contribute to the $\pi_0$ self energy
corrections, $\Sigma_{\pi^{0}}(\beta)$]{Diagrams that contribute to
the $\pi_0$ self energy corrections,
$\Sigma_{\pi^{0}}(\beta)$}\label{diagramas piones}
\end{center}
\end{figure}

\noindent whereas for the charged pions we have only the two
diagrams shown in Fig.5.

\begin{figure}[h]
\begin{center}
\includegraphics[scale=0.5]{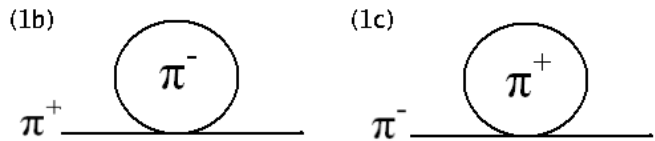}
\caption[Diagrams that contribute to the $\pi^{+}$, $\pi^{-}$ self
energy correction, $\Sigma_{\pi^{+}}(\beta,\mu_I)$ and
$\Sigma_{\pi^{-}}(\beta,\mu_I)$]{Diagrams that contribute to the
$\pi^{+}$, $\pi^{-}$ self energy correction,
$\Sigma_{\pi^{+}}(\beta,\mu_I)$ and $\Sigma_{\pi^{-}}(\beta,\mu_I)$}
\end{center}
\end{figure}

The self energy corrections $\sum$ allow, according to the well
known procedure of summing geometric series, to define a dressed
mass i.e. $m_{d}^2=m^{2}+i\sum$.

$\sum$ can be decomposed as $\sum=\sum(0)+\sum(\beta)$, where
$\sum(\beta)$ denotes the thermal corrections to the self energy.
$\sum(0)$ is the zero temperature self energy and it will not be
considered here, since it is absorbed according the usual
renormalization procedure at zero temperature.

The self energy correction for $\pi_0$ does not depend on the
isospin chemical potential $\mu_I$, since as we said we are not
considering here a $\mu_I$ associated to the fermions. On the
contrary, the self energy corrections to the charged pions will have
now thermal and $\mu_I$ dependent contributions.

Our results are plotted in Fig.6 for $\pi_0$ and in Fig.7 for the
charged pions. The chemical potential contributes, as it was
expected to increase de mass for the charged pions.

\begin{figure}[h]
\begin{center}
\includegraphics[scale=0.37]{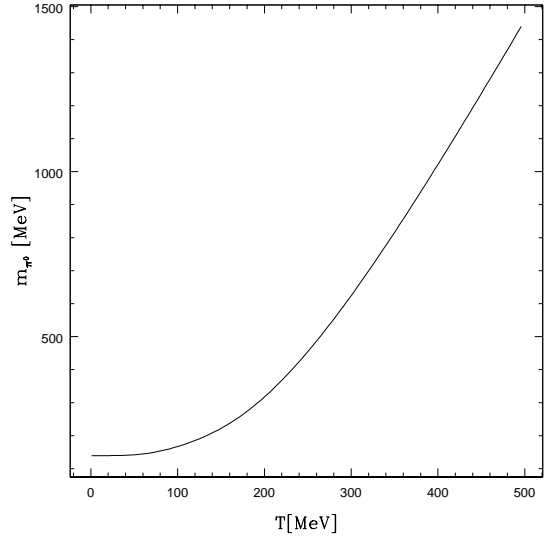}
\caption[Mass behavior of $m_{\pi_{0}}$.]{Mass behavior of
$m_{\pi_{0}}$.} \label{masapio}
\end{center}
\end{figure}

\begin{figure}[h]
\begin{center}
\includegraphics[scale=0.4]{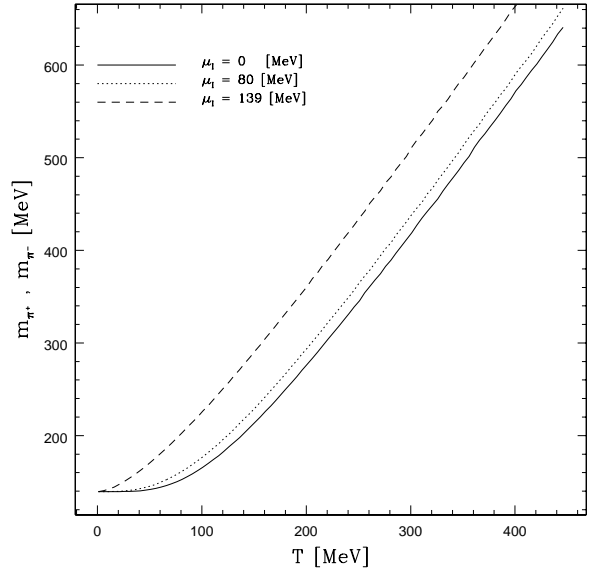}
\caption[Mass behavior of $m_{{\pi}^{+}}$ and $m_{{\pi}^{-}}$.
]{Mass behavior of $m_{{\pi}^{+}}$ and $m_{{\pi}^{-}}$.}
\label{masapimasmu}
\end{center}
\end{figure}

\pagebreak

%***************************************************************************************************
\section{$\sigma$ meson mass corrections}
%***************************************************************************************************

The relevant diagrams, where the $\sigma$ meson is represented by
the dashed line, are shown in Fig.8.

\begin{figure}[h]
\begin{center}
\includegraphics[scale=0.5]{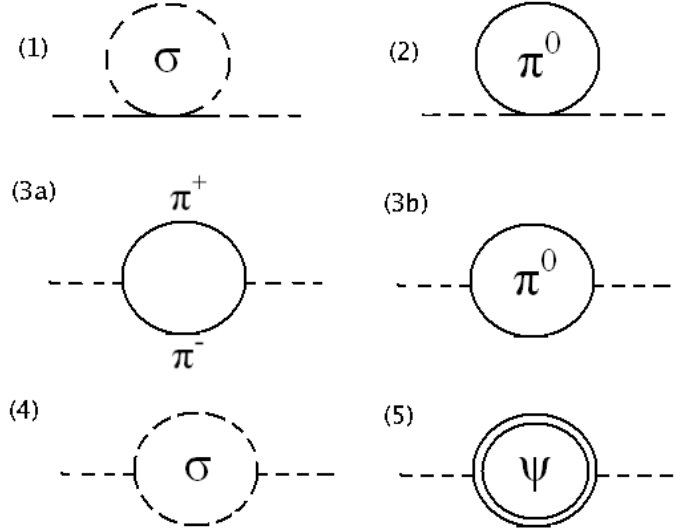}
\caption[Diagrams that contribute to
$\Sigma_{\sigma}(\beta)$]{Diagrams that contribute to
$\Sigma_{\sigma}(\beta)$}\label{diagramas sigma}
\end{center}
\end{figure}

The results for the sigma meson mass are shown in Fig.9 and Fig.10.
From the technical point of view, as usual, the principal value has
to be use in order to handle the poles.

The growing of $m_\sigma$, as function of T, does not correspond to
the $m_\sigma$ evolution in \cite{larsen} and in \cite{nico}. This
is due to the Mean Field approximation they followed. In fact, this
effective approach is quite more complicated than the usual loop
expansion for the masses.

\begin{figure}[h]
\begin{center}
\includegraphics[scale=0.4]{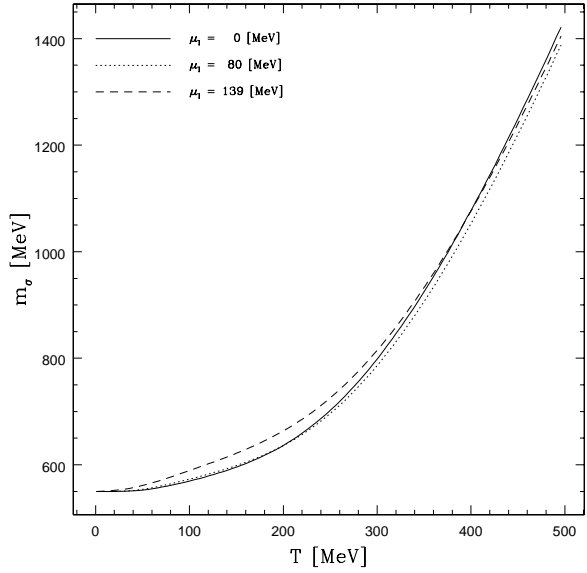}
\caption[Behavior of $m_{\sigma}$ in function of T. ] {Behavior of
$m_{\sigma}$ in function of T. } \label{masasigma}
\end{center}
\end{figure}

\begin{figure}[h]
\begin{center}
\includegraphics[scale=0.4]{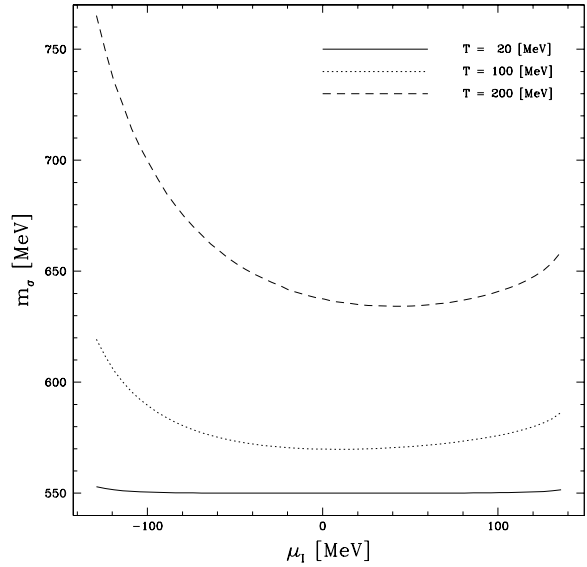}
\caption[Behavior of $m_{\sigma}$ in function of $\mu_I$. ]
{Behavior of $m_{\sigma}$ in function of $\mu_I$. }
\label{masasigma}
\end{center}
\end{figure}

 To summarize, we can see that the isospin chemical
potential contribution, both for the masses as well as for the
vacuum evolution is quite small compared to the temperature
dependence. The pion masses at the one loop order have have a very
similar behavior to the corresponding masses in the mean field
approaches. However, the sigma meson mass is quite different in both
approaches. The loop expansion predicts a steadily growing for the
sigma meson mass as function of T with small correction due to
$\mu_I$.

\section*{ACKNOWLEDGMENTS}
 The authors acknowledge support from
 FONDECYT under grant 1051067. M.L. also acknowledges support from the
 \textit{Centro de Estudios Subat\'omicos}. The authors would
 like to thank Dr. C. Contreras for helpful discussions. M.L.E.
 acknowledges also support from FONDECYT under grant 1060629.

\end{document}